\documentclass[useAMS,usenatbib,usegraphicx]{mn2e}

\newcommand{\cm}{cm$^{-1}$}
\newcommand{\teff}{T$_{\rm eff}$}

\newcommand{\AAA}{A\&A}
\newcommand{\AAAS}{A\&AS}
\newcommand{\ApJ}{ApJ}

\newcommand{\ApJS}{ApJSS}

\newcommand{\msol}{M$_\odot$}
\newcommand{\htp}{H$_3^+$}
\newcommand{\gcc}{g~cm$^{-3}$}
\newcommand{\saumon}{SBLHB94}
\newcommand{\marcs}{{\scshape marcs}}
\newcommand{\cesam}{{\scshape cesam}}
\newcommand{\ngelms}{{\scshape ng-elms}}

\title[Low mass primordial stellar evolution.]{Non-grey hydrogen burning evolution of sub-Solar mass Population III stars.}

\author[Harris et al.]{G. J. Harris$^1$, A. E. Lynas-Gray$^2$, S. Miller$^1$ \& J. Tennyson$^1$. \\
$^1$ Department of Physics \& Astronomy, University College London, London, WC1E 6BT, UK. \\
$^2$ Department of Physics, University of Oxford, Keble Road, Oxford OX1 3RH, UK.}

\begin{document} 

\maketitle
                                         
\begin{abstract}
The primordial elements, H, He and Li are included in a low temperature equation of state and monochromatic opacity function. The equation of state and opacity function are incorporated into the stellar evolution code \ngelms, which makes use of a non-grey model atmosphere computed at runtime. \ngelms\  is used to compute stellar evolution models for primordial and lithium free element mixtures, for stars in the sub-Solar mass range 0.8--0.15 \msol. We find that lithium has little or no effect upon the structure and observable properties of of stars in this mass range. Furthermore lithium is completely destroyed by fusion before the main sequence in stars of mass less than $\sim$0.7 \msol. We find that on the red giant branch and Hayashi track, the use of a non-grey model atmosphere to provide the upper boundary conditions for the stellar evolution calculation, results in significantly cooler less luminous stars, across the mass range. 

\end{abstract}


\section{Introduction.}
\label{sec:Intro}

The observation and identification of the first generation of stars (Population III) remains one of the main goals of modern astronomy. Indeed, \citet{Kashlinsky04} have investigated the possibility that light from the first stars may contribute to the near infrared background. They also suggest that this light may have already been observed via the Spitzer space telescope \citep{Kashlinsky06}. However \citet{Madau06} cast doubt upon this conclusion. 

If population III stars formed with masses less than 0.8 \msol\  they would still be on the main sequence at the present day. This leads to the possibility that these Population III stars may be present and observable within the solar neighbourhood. \citet{Christlieb} and \citet{Frebel} have reported the discovery of two low mass stars with [Fe/H] $<$ --5. It has been suggested that these stars may be population III stars which have accreated metals either from the interstellar medium of from a companion star to achieve their present day metallicities.

In the absence of metals most current star formation models predict that the Population III initial mass function is skewed to the very high mass ($>$ 100 M$_\odot$) \citep*{Brom99,Brom02,Abel03}. However this scenario is not yet certain, \citet{Tsujimoto} suggest that intermediate mass (3.5 -- 5 \msol) primordial stars are required to produce the heavy element abundance ratios seen in some metal poor stars. Furthermore, \citet{Nagakura} have proposed a method by which low mass primordial stars can form from the unpolluted remnants of high mass star formation, if the high mass star forms a black hole without supernova.

In our earlier paper \citep{Harris04a}, we computed evolutionary models of low-mass metal-free stars, using only a hydrogen and helium equation of state. It was demonstrated that \htp\  has a strong effect upon the effective temperature and luminosity of stars with M$<$0.4 \msol. However, these models were computed using grey model atmospheres to provide the outer boundary condition for the equations of stellar structure. For very low masses (M$<$0.5 \msol) the grey approximation is known to be inaccurate \citep{Saumon94,Baraffe96,Baraffe97,Chabrier97}. Additionally, our models neglected lithium, \citet{Mayer} have shown that Li can act as a significant electron donor at low temperatures in a primordial (H, He and Li) mixture. In this work we update our equation of state to include Li, and compute evolution models with a non-grey model atmosphere, from protostellar collapse to the Helium flash.

\section{Non-grey evolution.}
\label{sec:model}

The non-grey evolution of low mass stars (\ngelms) code is based upon a combination of 2 existing codes, \cesam\  \citep{Morel} and \marcs\  \citep{Gustafsson}. \cesam\  is a one dimensional stellar evolution code developed for the modelling of low and intermediate mass stars, which makes use of a grey model atmosphere. \marcs\  is a plane parallel model atmosphere code, for which we have rewritten the opacity and equation of state subroutines and modified to run as a subroutine of \cesam. \marcs\  provides the surface boundary conditions for \cesam\  at runtime. This code will be reported in more detail in a future article.

To provide nuclear reaction rates we use the NACRE \citep{NACRE} compilation, for the PP chain, CNO cycle and triple alpha process. The following reactions are accounted for: H(p,e$^+$~$\nu$)D, D(p,$\gamma$)$^3$He, $^3$He($^3$He,2p)$^4$He, $^4$He($^3$He,$\gamma$)$^7$Be,  $^7$Li(p,$\alpha$)$^4$He, $^7$Be(e$^-$,$\nu$~$\gamma$)$^7$Li, $^7$Be(p,$\gamma$)$^8$B(e$^+$~$\nu$)$^8$Be($\alpha$)$^4$He, $^{12}$C(p,$\gamma$)$^{13}$N(e$^+$~$\nu$)$^{13}$C, $^{13}$C(p,$\gamma$)$^{14}$N, $^{14}$N(p,$\gamma$)$^{15}$O(e$^+$~$\nu$)$^{15}$N, $^{15}$N(p,$\gamma$)$^{16}$O, $^{15}$N(p,$\alpha$)$^{12}$C, $^{16}$O($\alpha$,$\gamma$)$^{17}$F(e$^+$~$\nu$)$^{17}$O, $^{17}$O(p,$\alpha$)$^{14}$N, $^4$He(2$\alpha$,$\gamma$)$^{12}$C, $^{12}$C($\alpha$,$\gamma$)$^{16}$O, $^{16}$O($\alpha$,$\gamma$)$^{20}$Ne.

At low temperatures (T$<$9000~K) we use the equation of state and opacity function discussed in sections \ref{sec:LoMES} and \ref{sec:opac}. A linear interpolation upon the the OPAL equation of state \citep{OPALEoS} and opacity  \citep{OPALop} tables provides the equation of state and opacity data at higher temperatures (T $>$ 11~000~K). Between temperatures of 9000 and 11~000~K a linear interpolation between OPAL data and the data of this work is used. The OPAL equation of state tables, cover the temperatures and densities of stars with masses greater than about 0.15 \msol. This limits our calculation of evolution models to masses of 0.15 \msol\  and greater. Conductive opacities are obtained by interpolating the tables of \citet{condop}. The diffusion of elements including metals is accounted for via the method of \citet{Michaud}.
The mixing length theory of convection is used, with an adopted value of the mixing length parameter of $\alpha=1.745$. This value of the mixing length parameter was achieved via solar calibration, using an equation of state which contains the elements H, He, Li, C, N, O, Ne, Na, Mg, Al, Si, S, Ca, Ti, Fe; this will be fully reported in a future article. At the accepted solar age of 4.5 Gyr, for a solar mass model we achieve \teff$=5781.8$, $L/L_\odot=1.00398$ and $R/R_\odot=1.000497$. Our initial  mass fractions are $X_0=0.70440$, $Z_0=0.018698$, and surface mass fractions at the solar age are $X_s=0.72889$, $Z_s=0.017783$, giving $Z_s/X_s=0.02440$, this is comparable with the solar value of $Z_\odot/X_\odot=0.02440$ \citep{Grevesse}. These initial conditions compare with the solar evolution and helioseismology model fits of \citet{Guenther}, who find $Z_0$=0.020 and $X_0=0.7059$, and achieve $Z_\odot/X_\odot=$0.0244. 

During the course of this work \citet*{Asplund} have published new estimates of the metal abundance ratios and metallicity of the Sun. They revise downward the metal to hydrogen mass fraction ratio to $Z/X=0.0165$. As metals are important contributors to the solar opacity, the reduced metallicity will result in a lower opacity. So that a solar model computed with the new metal mix will have a greater effective temperature, luminosity and radius than one computed with the metal mix of \citep{Grevesse}. To counter this, the mixing length parameter will need to be revised downward.

\subsection{Low temperature equation of state.}
\label{sec:LoMES}


In order to compute opacities it is first necessary to know the abundances of each atomic molecular and ionic species. The equation of state computes the various thermodynamic quantities as well as the number densities of each species. Our low temperature equation of state ({\scshape lomes}) is an extensively rewritten version of the equation of state used in \citet{Harris04a}. This equation of states has been used by \citet{Engel} and in a modified, non-ideal, form by \citet{Harris04b} to demonstrate the importance of HeH$^+$ at very low hydrogen abundances. In light of the findings of \citet{Mayer}, we have updated the equation of state to include Li, Li$^+$, Li$^{++}$, LiH and LiH$^+$. {\scshape lomes} is based upon the Saha equation. Each species is written in terms of its constituent neutral atomic species and electrons. For example the formation reaction of \htp\  is written as 3H$\rightleftharpoons$\htp+$e^-$, the Saha equation is then
\begin{equation}
\frac{N(H_3^+)N(e^-)}{[N(H)]^3}=\frac{Q_T(H_3^+)Q_T(e^-)}{[Q_T(H)]^3}\exp\left(-\frac{\Delta E}{kT}\right)
\end{equation}
where $N$ are the number densities of each species, $Q_T$ are the total partition functions, $\Delta E$ is the energy difference between the left and right hand sides of the reaction equation. The Saha equation for each species is used to construct simultaneous equations for the conservation of charge and conservation of nucleons, which are then solved. This is a widely used technique, see for example \citet{Kurucz1970} and \citet{Mayer}. The rewritten algorithm has none of the convergence problems reported in \citet{Harris04a}. The primordial atomic, molecular and ionic species included in the equation of state are quoted in table \ref{tab:partfunc}. For the mass range covered here, the densities of the regions below 9000~K are not sufficiently high to result in strong non-ideal effects such as pressure dissociation and pressure ionisation. Thus a Saha based equation of state is adequate for the purposes of this study.

\begin{table*}
 \centering
 \begin{minipage}{140mm}
\caption{The primordial atomic, molecular and ionic species included in the equation of state and the source of the partition function used.}
\begin{tabular}{lcl}
\hline
Species                    & T range (K) & Reference \\ \hline
  H$_2$                    & 1000--9000  & \citet{Sauval} \\
  H                        & 1000--16000 & \citet{Irwin} \\
  H$^-$                    &  -          &  \\
  H$^+$                    &  -          &  \\
  H$_2^+$                  & 1500--18000 & \citet{Stancil} \\
  H$_2^-$                  & 1000--9000  & \citet{Sauval} \\
  \htp                     & 500--8000   & \citet{Neale95} \\
  He                       & 1000--16000 & \citet{Irwin} \\
  He$^+$                   & 1000--16000 & \citet{Irwin} \\
  HeH$^+$                  & 500--10000  & \citet{Engel} \\
  Li                       & 1000--16000 & \citet{Irwin} \\
  Li$^+$                   & 1000--16000 & \citet{Irwin} \\
  Li$^++$                  & 1000--16000 & \citet{Irwin} \\
  LiH                      & 200--20000  & \citet{Stancil}\\
  LiH$^+$                  & 200--20000  & \citet{Stancil}\\
  e$^-$                    &       -     & -  \\ \hline
\end{tabular}
                                                                                
\label{tab:partfunc}
\end{minipage}
\end{table*}                                                                               

\begin{figure*}
\includegraphics[angle=-90,width=80mm]{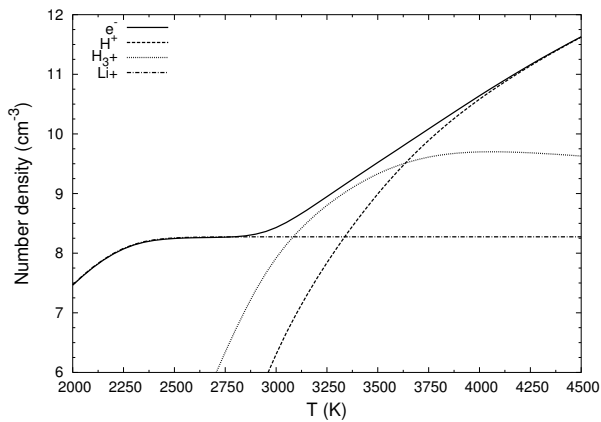}
\caption{The number densities of the dominant positive ions and electrons at a density of 10$^{-6}$ \gcc.}
\label{fig:EoS}
\end{figure*}

Figure \ref{fig:EoS} shows how number density of the dominant positive ions and electrons change with temperature for a density of 10$^{-6}$ \gcc. As found by \citet{Mayer} ionised lithium is the dominant positive ion at low temperatures, above 3000 K the abundance of \htp\  exceeds that of Li$^+$ and H$^+$ becomes the dominant positive ion for temperatures greater than 3600 K. At densities lower than $10^{-7}$ \gcc, the abundance of \htp\  is hugely reduced by free electrons from Li, to the extent at which it becomes unimportant. At these low densities, Li$^+$ is the dominant positive ion at low temperatures and H$^+$ is dominant at high temperatures.

\subsection{Low temperature opacities}
\label{sec:opac}

Our sources of continuous opacity and scattering data are the same as those used in \citet{Harris04a}. Most of our computations involve non-grey atmospheres which require monochromatic opacities over a range of frequencies. The use of line opacity in the computation of model atmospheres is problematic \citep{Carbon}, by its nature line opacity is not smooth. Spectroscopic lines block electromagnetic flux from escaping an atmosphere near the line centres, but allow flux to escape between lines. Much work has gone into statistical treatments of line opacity within model atmospheres, which has resulted in some success for opacity sampling and the opacity distribution function (ODF), see for example \citet{Ekberg}. In \ngelms\  it is necessary for model atmospheres to be computed approximately 30 times per evolutionary time step, it is therefore essential for a solution to be arrived at rapidly. With this in mind we use a straight mean opacity computed over the interval between frequency points, for each source of line opacity. \citet{Carbon} has shown that the straight mean tends to over estimate opacity, we find that the systematic errors introduced by a straight mean are relatively small from many closely packed weak lines, such as molecular bands. However, for strong isolated lines, such as atomic lines, the opacity is dramatically over predicted. For the mass range studied here the atomic (H, Li) and molecular (H$_2$) lines make up a relatively small fraction of the total opacity, we have therefore neglected H$_2$, Li and H lines from our calculations. To test if line opacity from H$_2$, Li and H, affects the evolution of low mass stars we have computed ODFs for these species. Line lists for H and Li are taken from \citet{Kurucz1993}. Einstein A coefficients for H$_2$ quadrupole transitions taken from \citet{Wolniewicz} and H$_2$ energy levels are taken from the Molecular Opacity Database UGAMOP (http://www.physast.uga.edu/ugamop/). Two evolutionary models are computed using these ODFs and are discussed in section \ref{sec:results}.

\begin{figure*}
\includegraphics[angle=-90,width=80mm]{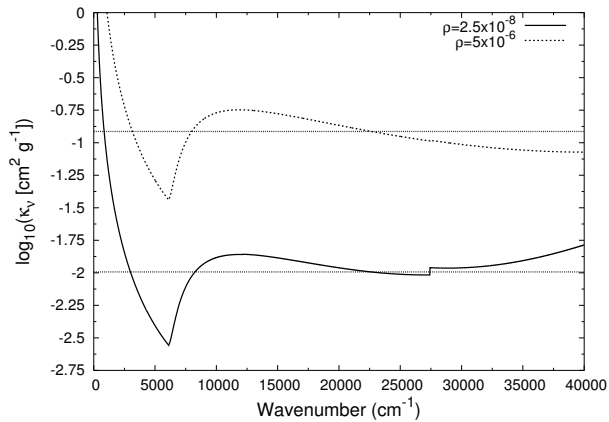}
\caption{Continuous opacity at 5000~K, for densities of 2.5$\times10^{-8}$ and $5.0\times10^{-6}$ \gcc, which are the approximate densities found in giant and dwarf photosphere respectively. For reference, the Rosseland mean opacity is shown as a horizontal line for each of the two curves.}
\label{fig:opac}
\end{figure*}

Figure \ref{fig:opac}, shows the continuous opacity and Rosseland mean opacity at 5000~K for densities of 2.5$\times10^{-8}$ and $5.0\times10^{-6}$ \gcc. For the higher density (5$\times10^{-6}$ \gcc) opacity is completely dominated by H$^-$ bound-free and free-free continuous opacity. The continuous opacity is slightly less than the Rosseland mean for frequencies above 22~000 \cm, and slightly greater than the Rosseland mean for frequencies between 8000 and 22~000 \cm. Thus, overall the Rosseland mean opacity, which is a weighted harmonic mean, gives a reasonable representation of opacity over the frequency range of the peak of the Planck function. At the lower density (2.5$\times10^{-8}$ \gcc), the opacity is still dominated by H$^-$ in the red, but in the blue the contribution to opacity from Rayleigh scattering and H bound-free continuous opacity are important. The Balmer jump is noticeable and the Rosseland mean opacity appears to be too low to provide a good representation of overall opacity. As is discussed in section \ref{sec:atmos}, this 'overestimate' of opacity leads to the breakdown of the grey approximation for giant stars.

Radiative transport in the interior of the star is treated using the diffusion approximation, which requires the Rosseland mean opacity. Our grey model atmospheres also use the Rosseland mean opacity. The Rosseland mean opacity is computed at runtime, via numerical integration using 145 frequency points. 

\subsection{Grey and non-grey model atmospheres.}
\label{sec:atmos}

\begin{figure*}
\includegraphics[angle=-90,width=80mm]{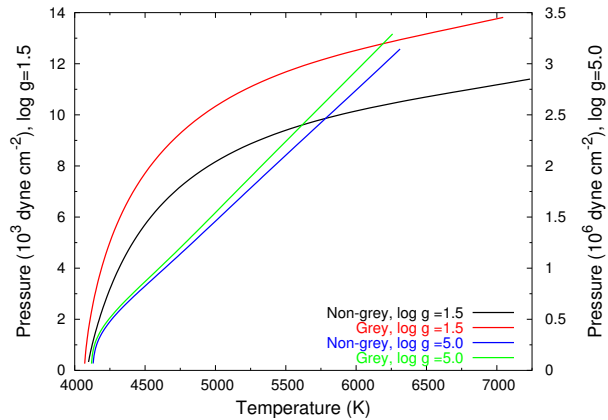}
\caption{Temperature pressure curves for grey and non-grey model atmospheres at \teff=5000~K, and with $\log g=1.5$ and 5.0. Pressure for the lower surface gravity is given on the left axis and for the higher gravity denser atmosphere on the right axis.}
\label{fig:atmos}
\end{figure*}

It has previously been found that for dwarf stars with \teff\  less than $\sim$5000~K, the grey approximation provides inaccurate boundary conditions for stellar evolution calculations \citep{Saumon94,Baraffe96,Baraffe97,Chabrier97}. Figure \ref{fig:atmos} shows the temperature-pressure curves of 4 grey and non-grey model atmospheres, computed for \teff=5000~K and with surface gravities of $\log g=1.5$ (giant) and $\log g=5$ (dwarf). It is clear that for both surface gravities the grey approximation provides a poor representation of the non-grey atmosphere. In general the non-grey atmosphere is less dense than the grey atmosphere, this is consistent with the grey atmosphere having an effectively lower opacity than the non-grey atmosphere. There is a far stronger deviation of grey atmosphere from non-grey atmosphere at low surface gravities. 

As temperature is reduced the non-grey effect strengthens. For \teff\  above $\sim$6000~K the grey atmosphere gives good agreement with the non-grey atmosphere, even for low surface gravities. Non-grey model atmospheres must be used to provide the boundary conditions for stellar evolution models below 6000~K, particularly for giants.

\subsection{Computation of evolution models.}
\label{sec:models}


In the computation of evolution models we use two separate element mixtures. A lithium free mixture which contains only hydrogen and helium and a primordial mixture which in addition to H and He also contains Li. We adopt the element abundance ratios determined by \citet{Coc} from big bang nucleosynthysis model fits to the cosmic microwave background observations by the Wilkinson microwave anisotropy probe (WMAP). It must be noted that the value of the Li/H ratio of $4.25^{+0.49}_{-0.45}\times10^{-10}$, determined by \citet{Coc}, deviates by a factor of 3 from the value of $1.23^{+0.68}_{-0.32}\times10^{-10}$ determined from observation of halo stars \citep{Ryan}. We adopt the higher value in order to asses the maximum impact of Li on low mass stellar evolution in the early Universe. The adopted element abundances for the primordial mix is $X=0.75210$ and $Z_{\rm Li}=2.24\times10^{-9}$, and for the lithium free mix is $X=0.75210$ and $Z_{\rm Li}=0$.

All Models are computed from an initial quasi-static protostellar model undergoing gravitational collapse. As in our earlier work \citep{Harris04a} we use a contraction constant of 0.015 L$_\odot$M$^{-1}_\odot$K$^{-1}$. The models are followed through collapse down the Hayashi track, through the main sequence. Models are allowed to evolve onto and up the red giant branch to near the point of helium flash, or the point at which stars of insufficient mass to fuse Li turn off the red giant branch. With \ngelms\  we have computed evolutionary models for stars with masses of between 0.15 and 0.8 \msol, using grey and non-grey model atmospheres, for both primordial and Li free mixtures.


\section{Results.}
\label{sec:results}

\begin{figure*}
\includegraphics[angle=-90,width=80mm]{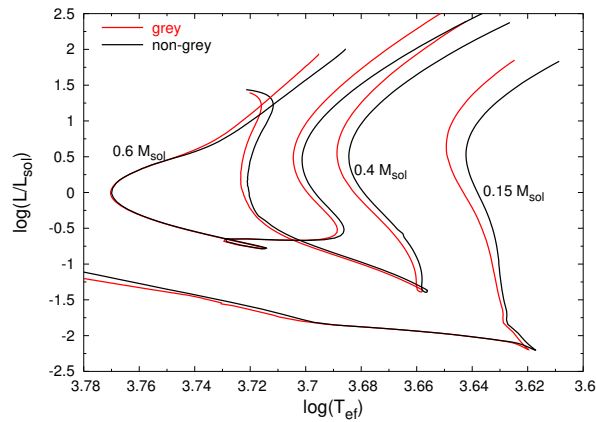}
\caption{Evolutionary tracks of stars of masses 0.15, 0.4, and 0.6 \msol, computed with and without a grey model atmosphere.}
\label{fig:HRgrey}
\end{figure*}

Evolutionary tracks on a HR diagram, for stars of, 0.15, 0.4, and 0.6 \msol, computed using grey and non-grey model atmospheres, are shown in figure \ref{fig:HRgrey}. As we have previously found \citep{Harris06}, on the red giant branch and Hayashi track, all the non-grey models have significantly lower effective temperatures for a given luminosity than their counterparts computed with grey model atmospheres. Furthermore, on the Main Sequence the non-grey models with masses less than $\sim$0.5 \msol, are also cooler than their grey counterparts, which is consistent with the findings of \citet{Saumon94}, hereafter \saumon, and of \citet{Baraffe96,Baraffe97,Chabrier97}. These two factors again, illustrate the importance of using a non-grey model atmosphere to provide the boundary conditions for evolution.

\begin{figure*}
\includegraphics[angle=-90,width=80mm]{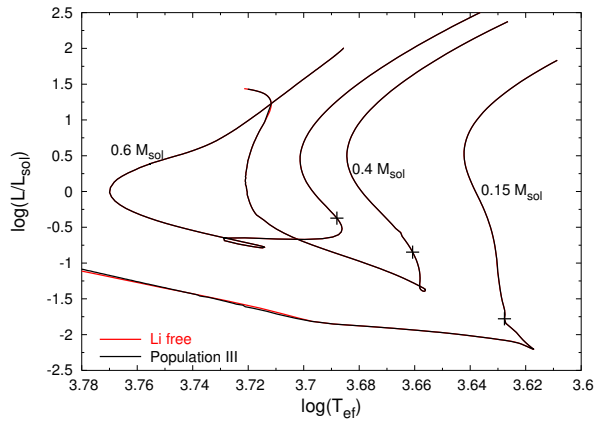}
\caption{Evolutionary tracks of stars of masses 0.15, 0.4, and 0.6 \msol, with both primordial and lithium free element mixtures. Crosses on the evolutionary tracks mark the point at which Li fusion ends.}
\label{fig:HRLi}
\end{figure*}

 Figure \ref{fig:HRLi} shows the evolutionary tracks of models of 0.15, 0.4, and 0.6 \msol, computed with primordial and lithium free element mixtures. We find that Li has little or no effect upon the evolution of stars of mass between 0.8 and 0.15 \msol. The reason for this is that all the stars are hot enough to either ionise hydrogen or form \htp, releasing more electrons than can be produced by the ionisation of Li. Furthermore, lithium is destroyed within a few million years by fusion via the reaction $^7$Li(p,$\alpha$)$^4$He, which occurs before the  star reaches the Main Sequence. As primordial stars with mass less than $\sim$0.5 \msol\  are fully convective at zero age Main Sequence, all the lithium in the cooler lower mass stars is destroyed. 

\begin{figure*}
\includegraphics[angle=-90,width=80mm]{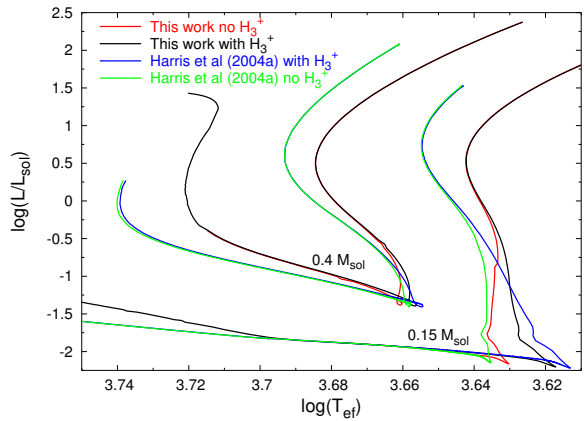}
\caption{Evolutionary tracks of stars of masses 0.15 and 0.4 \msol, computed with and without \htp\  included in the equation of state. Also shown are the evolutionary tracks of \citet{Harris04a}.}
\label{fig:HRH3p}
\end{figure*}

In our earlier work, \citet{Harris04a} we demonstrated the important role that \htp\  has upon the evolution of stars of mass less than 0.4\msol. In order to re-asses the impact of \htp\  in light of the non-grey model atmospheres we have computed models with and without \htp\  in the equation of state. Figure \ref{fig:HRH3p}, shows the evolutionary tracks of models of masses 0.15 and 0.4 \msol\  computed with and without \htp\  in the equation of state. Also plotted are our previous models \citep{Harris04a} which are computed with a grey atmosphere and with the higher hydrogen mass fraction of $X=0.77$. As the rate of the reaction H(p,e$^+$~$\nu$)D, is proportional to the square of the abundance of hydrogen, the higher mass fraction results in slightly different positions of the zero age main sequence on the HR diagram. Another difference is that \citet{Harris04a}, used the slightly higher value of the mixing length parameter $\alpha=0.18$. This is important for the pre-Main and post-Main Sequence evolution, but has only small impact upon the Main Sequence. 

\begin{figure*}
\includegraphics[angle=-90,width=80mm]{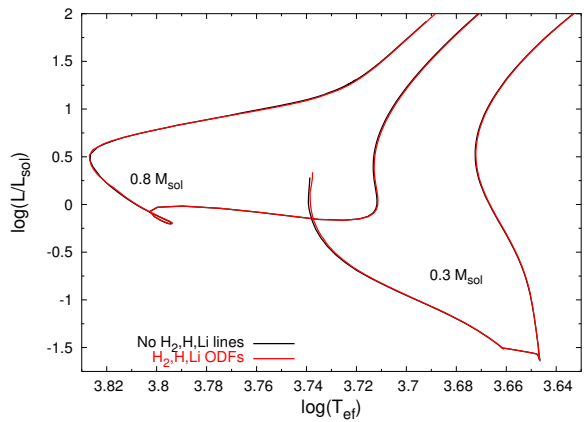}
\caption{Evolutionary tracks for primordial element composition stars of masses 0.8 and 0.3 \msol, computed with and without ODFs to account for the line opacity of H, H$_2$ and Li.}
\label{fig:odf}
\end{figure*}

Figure \ref{fig:odf} show evolutionary tracks computed with primordial abundances, for 0.8 and 0.3 \msol\  stars. One set of models uses ODFs, described in section \ref{sec:opac}, to represent opacity from H$_2$, H and Li lines and the other neglects line opacity for these species. There is almost no difference between the 0.8 and 0.3 \msol\  evolutionary tracks computed with the H, H$_2$ and Li ODFs and neglecting H, H$_2$ and Li opacity. So, H$_2$, H and Li lines do not have any significant affect on the evolutionary models presented here.

\citet{Siess} have computed zero metallicity evolution models for the mass range 0.8 to 20 \msol, and \citet{Marigo} for 0.7 to 100 \msol. The core hydrogen burning evolutionary lifetimes given by \citet{Siess} and \citet{Marigo} for a 0.8 \msol model are 14.8 and 13.75 Gyr respectively. This compares with our computed hydrogen burning lifetime of 11.96 Gyr. The 2-3 Gyr difference between hydrogen burning lifetimes computed in this work and by \citet{Siess} and \citet{Marigo} is primarily due to the use of different initial hydrogen mass fractions. \citet{Siess} and \citet{Marigo} use an initial H mass fraction of 0.77, compared to the value of 0.752 used in this work.

\begin{figure*}
\includegraphics[angle=-90,width=80mm]{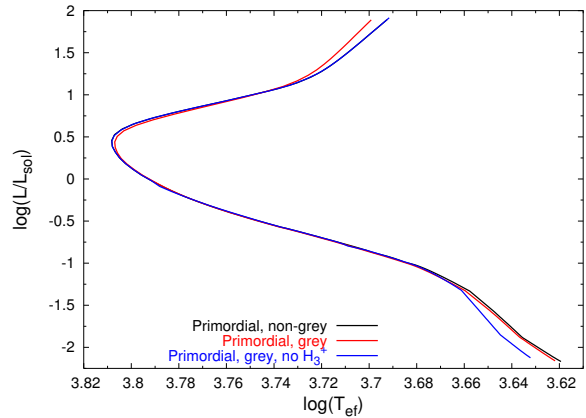}
\caption{14 Gyr isochrone, determined from models computed with primordial element abundances and with grey and non-grey atmospheres, and a grey model atmosphere with \htp\  neglected from the equation of state.}
\label{fig:isoc-giant}
\end{figure*}

\begin{figure*}
\includegraphics[angle=-90,width=80mm]{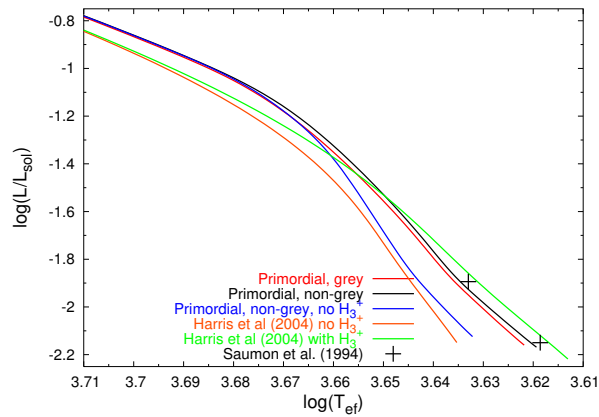}
\caption{14 Gyr isochrone, determined from models computed with primordial element abundances and with grey and non-grey atmospheres and without \htp. Also shown are the 14 Gyr isochrones of \citet{Harris04a}, and the 0.15 and 0.2 \msol\  models of \saumon.}
\label{fig:isoc}
\end{figure*}

Figure \ref{fig:isoc-giant} shows the 14 Gyr isochrone derived from our models computed using primordial abundances, with grey and non-grey atmospheres and with and without \htp. The primary differences between the grey and non-grey isochrones occur upon the red giant branch, where models computed with non-grey boundary conditions result in lower effective temperatures for a given luminosity. Figure \ref{fig:isoc} shows the low mass end of the 14 Gyr isochrone computed with grey and non-grey model atmospheres and with and without \htp. Also shown in this figure are the isochrones of \citet{Harris04a} and the models of \saumon.
 The addition of \htp\  clearly results in a significant drop in \teff\, the use of the grey model atmosphere also decreases \teff. The models of \saumon\  fall between the non-grey isochrone of this work and the grey isochrone of \citet{Harris04a}.Table \ref{tab:colour} lists broad band colours and atmospheric parameters, for our non-grey evolutionary models. For masses below 0.5 \msol, the models which exclude \htp\  are hotter, show bluer colours, and are correspondingly more luminous than those which account for \htp. There is in general good agreement between the colours of \saumon\  and those computed in this work, although the atmospheres and colours of \saumon\  are slightly cooler and redder. This may be due to the fact that \saumon\  use older collision induced absorption data, which in general gives a higher opacity and may result in a cooler star. The broadband color indices were computed by using the bandpasses given by \citet{Bessell88,Bessell90} and were calibrated by using a spectrum of Vega.

\begin{table*}
 \centering
 \begin{minipage}{140mm}
\caption{Broad band colours and atmospheric parameters of 14 Gyr primordial non-grey model stars, computed with and without \htp\  in the equation of state. Also listed is the data of \saumon.}
\begin{tabular}{clllccccccccc}
\hline
M(\msol) & Source  & \htp & \teff  & $\log_{10}(L/L_\odot)$ & $\log_{10} g$ & B-V  & V-R  & V-I  & V-J  & V-H  & V-K  \\ \hline

0.75 & This work   & yes  & 6342.9 &  0.501                 & 3.97          & 0.41 & 0.29 & 0.60 & 0.93 & 1.20 & 1.24  \\
   & This work   & no  & 6342.9 &  0.501                 & 3.97          & 0.41 & 0.29 & 0.60 & 0.93 & 1.20 & 1.24  \\
\\
0.70 & This work   & yes  & 6188.3 & -0.020                 & 4.42          & 0.45 & 0.31 & 0.64 & 0.99 & 1.28 & 1.32 \\
     & This work   & no   & 6188.3 & -0.020                 & 4.42          & 0.45 & 0.31 & 0.64 & 0.99 & 1.28 & 1.32 \\
\\
0.60 & This work   & yes  & 5459.5 & -0.571                 & 4.69          & 0.61 & 0.41 & 0.83 & 1.29 & 1.66 & 1.74 \\
     & This work   & no   & 5458.2 & -0.572                 & 4.69          & 0.61 & 0.41 & 0.83 & 1.30 & 1.66 & 1.74 \\
\\
0.50 & This work   & yes  & 4804.0 & -1.012                 & 4.83          & 0.81 & 0.53 & 1.06 & 1.65 & 2.06 & 2.21 \\
     & This work   & no   & 4810.0 & -1.011                 & 4.83          & 0.81 & 0.53 & 1.06 & 1.65 & 2.06 & 2.20 \\
\\
0.40 & This work   & yes  & 4547.0 & -1.332                 & 4.96          & 0.90 & 0.58 & 1.17 & 1.82 & 2.24 & 2.41 \\
     & This work   & no   & 4583.9 & -1.322                 & 4.96          & 0.89 & 0.58 & 1.15 & 1.79 & 2.20 & 2.37 \\
\\
0.30 & This work   & yes  & 4445.5 & -1.566                 & 5.03          & 0.93 & 0.60 & 1.21 & 1.89 & 2.31 & 2.47 \\
     & This work   & no   & 4517.9 & -1.543                 & 5.03          & 0.92 & 0.59 & 1.18 & 1.82 & 2.17 & 2.23 \\
\\
0.20 & This work   & yes  & 4320.0 & -1.885                 & 5.12          & 0.97 & 0.63 & 1.26 & 1.97 & 2.36 & 2.46 \\
     & This work   & no   & 4413.7 & -1.853                 & 5.12          & 0.95 & 0.61 & 1.21 & 1.87 & 2.18 & 2.18 \\
     & \saumon & - & 4298 & -1.896       & 5.12          & -    & 0.63 & 1.27 & 2.02 & 2.41 & 2.53 \\
\\
0.15 & This work   & yes  & 4162.6 & -2.169                 & 5.21         & 1.02 & 0.66 & 1.32 & 2.07 & 2.41 & 2.43 \\
     & This work   & no   & 4287.0 & -2.124                 & 5.22         & 0.98 & 0.63 & 1.25 & 1.92 & 2.14 & 2.09 \\
     & \saumon & - & 4155 &  -2.152      & 5.19          & -    & 0.66 & 1.33 & 2.10 & 2.43 & 2.51 \\ \hline
\end{tabular}
\label{tab:colour}
\end{minipage}
\end{table*}




\section{Conclusion.}
\label{sec:conc}

Lithium based species have been added to our equation of state, enabling us to confirm the finding of \citet{Mayer} that ionised lithium is the dominant positive ion at low temperatures, indirectly increasing opacity. This equation of state has been included in our opacity computations, which indicate that the Rosseland mean opacity is a poor representation of opacity at low densities and for temperatures below 6000~K. Our grey and non-grey model atmospheres also show that a grey atmosphere is inadequate for giant stars with \teff$<$ 6000~K. We also confirm the finding of \saumon that the Grey approximation is invalid for dwarf stars with \teff$<$5000~K. However, we find the deviation of grey from non-grey model atmosphere to be smaller than reported by \saumon, which may be due to the use of older collision induced absorption data by \saumon.

We have computed a set of stellar evolution models for primordial stars of mass 0.8--0.15 \msol, using a non-grey model atmosphere to provide the surface boundary conditions. The models are computed from protostellar collapse to the peak of the red giant branch or the point at which the star turns of the giant branch. These models have \teff\  high enough ($>4000$~K) to ensure sufficient electrons are released via the formation of H$^+$ or \htp, so that lithium has little or no effect on protostellar collapse and stellar evolution. Furthermore Li is fully destroyed before the Main Sequence in stars of mass less than $\sim$0.7 \msol. However, the role of lithium in the formation of primordial stars and low mass brown dwarfs, which are to cool to burn lithium, still needs to be investigated.

In line with earlier work \saumon\  we find that using a non-grey model atmosphere gives a lower effective temperature and luminosity for stars of mass less than 0.5\msol. However, we find the deviation of Main-Sequence grey from non-grey models is smaller than reported by \saumon, which again may be due to the use of older collision induced absorption data. More importantly, our non-grey models have significantly lower effective temperatures for a given luminosity, on the Hayashi track and red giant branch, than do our grey models. We also find that excluding \htp\  from the computations of stellar evolution for stars with mass less than 0.5 \msol\  results in cooler, bluer colours and less luminous stars.

\section*{Acknowledgements}

We thank the UK Particle Physics and Astronomy Research Council (PPARC) for funding.


\begin{thebibliography}{99}

\bibitem[\protect\citeauthoryear{Abel et al.}{2002}]{Abel03} Abel T., Bryan G. L., Norman M. L. 2002, Science, 295, 93.
\bibitem[\protect\citeauthoryear{Angulo et al.}{1999}]{NACRE}
Angulo, C., Arnould, M., Rayet, M., et al., 1999, Nuclear Physics A, 656, 3.
\bibitem[\protect\citeauthoryear{Asplund, Grevesse \& Sauval}{Asplund et al.}{2005}]{Asplund}
Asplund, M., Grevesse, N.,\& Sauval, A. J., 2005, in Cosmic Abundances as Records of Stellar Evolution, eds. Barnes III, T. G., Bash, F. N. (ASP conference series) vol. 336, P.\ 25
\bibitem[\protect\citeauthoryear{Carbon}{1974}]{Carbon}
Carbon D. F., 1974, \ApJ, 187, 135.
\bibitem[\protect\citeauthoryear{Baraffe \& Chabrier}{1996}]{Baraffe96}
Baraffe, I., \&  Chabrier, G. 1996, in from stars to galaxies, eds. C. Leitherer, U. Fritze-von-Alversleben, \& J. Huchra (ASP conference series) vol. 98, P.\ 209
\bibitem[\protect\citeauthoryear{Baraffe et al.}{1997}]{Baraffe97}
Baraffe, I., Chabrier, G., Allard, F., \& Hauschildt, P. 1997, A\&A, 327,1054.
\bibitem[\protect\citeauthoryear{Bessell \&\  Brett}{1988}]{Bessell88} 
Bessel, M. S., Brett, J. M., 1988, Pub. Astron. Soc. Pacific, 100, 1134.
\bibitem[\protect\citeauthoryear{Bessell}{1990}]{Bessell90} 
Bessel, M. S., 1990, Pub. Astron. Soc. Pacific, 102, 1181.
\bibitem[\protect\citeauthoryear{Borysow et al.}{2001}]{Borysow} Borysow, A., J\o rgensen, U. G., and Fu, Y. 2001, J. Quant. Spectrosc. Rad. Trans.,  68, 235.
\bibitem[\protect\citeauthoryear{Bromm, Coppi \& Larson}{Bromm et al.}{1999}]{Brom99} Bromm V., Coppi P. S., Larson R. B. 1999, ApJ, 527, L5 
\bibitem[\protect\citeauthoryear{Bromm, Coppi \& Larson}{Bromm et al.}{2002}]{Brom02} Bromm V., Coppi P. S., Larson R. B. 2002, ApJ, 564, 23
\bibitem[\protect\citeauthoryear{Chabrier \& Baraffe}{1997}]{Chabrier97}
Chabrier, G., Baraffem I. 1997, A\&A 327, 1039.
\bibitem[\protect\citeauthoryear{Coc et al.}{2004}]{Coc}
Coc A., Vangioni-Flam E., Descouvemont P., Adahchour A., Angulo C., 2004, \ApJ, 600, 544.
\bibitem[\protect\citeauthoryear{Ekberg, Eriksson \& Gustafsson}{Ekberg et al.}{1986}]{Ekberg} 
Ekberg U., Eriksson K. \& Gustafsson B., 1986, A\&A, 167, 304.
\bibitem[\protect\citeauthoryear{Engel et al.}{2005}]{Engel} Engel E. A., Doss N., Harris G. J., Tennyson J., 2005, MNRAS, 357, 471.
\bibitem[\protect\citeauthoryear{Christlieb et al.}{2002}]{Christlieb} Christlieb N., Bessell M. S. Beers T. C., Gustafsson B., Korn A., Barklem P. S., Karlsson T., Mizuno-Wiedner M., and Rossi S. 2002, Nature, 419, 904.
\bibitem[\protect\citeauthoryear{Frebel et al.}{2005}]{Frebel} Frebel A., Aoki W., Christlieb N., Ando H., Asplund M., Barklem P. S., Beers T. C., Eriksson K., et al., 2005, Nature, 434, 871.
\bibitem[\protect\citeauthoryear{Grevesse, Noels \& Sauval}{Grevesse et al.}{1996}]{Grevesse} Grevesse N., Noels A., \& Sauval A. J.,1996, in Cosmic abundances, eds Holt S. S., Sonneborn G., (ASP conference series), 117.
\bibitem[\protect\citeauthoryear{Guenther \& Demarque}{1997}]{Guenther} Guenther D. B.., Demarque, P., 1997, \ApJ, 484, 937
\bibitem[\protect\citeauthoryear{Gustafsson et al.}{1975}]{Gustafsson} Gustafsson B., Bell R. A., Eriksson K., Nordlund \AA., 1975, \AAA, 42, 407
\bibitem[\protect\citeauthoryear{Gustafsson \& Frommhold}{2001}]{Gustafsson01} Gustafsson M., and Frommhold L. 2001, \ApJ. 546, 1168. 
\bibitem[\protect\citeauthoryear{Harris et al.}{2006}]{Harris06}
 Harris G. J., Lynas-Gray, A.E., Tennyson, Miller S., 2006, in stellar evolution at low metallicity: mass loss, explosions and Cosmology, eds. Lamers H., Langer N., Nugis T., Annuk K., (ASP conference series)
\bibitem[\protect\citeauthoryear{Harris et al.}{2004a}]{Harris04a} Harris G. J., Lynas-Gray, A.E., Miller S., Tennyson J., 2004a, \ApJ, 600, 1025
\bibitem[\protect\citeauthoryear{Harris et al.}{2004b}]{Harris04b} Harris G. J., Lynas-Gray, A.E., Miller S., Tennyson J., 2004b, \ApJ, 617, L143
\bibitem[\protect\citeauthoryear{Hubbard \& Lampe}{1969}]{condop}
Hubbard, W. B., \& Lampe, M., 1969, ApJS, 18, 297.
\bibitem[\protect\citeauthoryear{Iglesias \& Rogers}{1996}]{OPALop}
Iglesias, C. A. and Rogers, F. R. 1996, ApJ, 464, 943.
\bibitem[\protect\citeauthoryear{Irwin}{1981}]{Irwin} Irwin A. W., 1981, \ApJS, 45, 621.
\bibitem[\protect\citeauthoryear{J\o rgensen et al.}{2000}]{Jorgensen} J\o rgensen U. G., Hammer D., Borysow A., and Falkesgaard J. 2000, \AAA, 361, 283.
\bibitem[\protect\citeauthoryear{Kashlinsky}{2006}]{Kashlinsky06} Kashlinksky A., IN PRESS, ApJ, astro-ph/0508089.
\bibitem[\protect\citeauthoryear{Kashlinsky et al.}{2004}]{Kashlinsky04} Kashlinksky A., Arendt R., Gardner J.P., Mather J.C., Moseley S. H., \ApJ, 608, 1.
\bibitem[\protect\citeauthoryear{Kurucz}{1993}]{Kurucz1993}
Kurucz, R.L., 1993, CD-ROM 1.
\bibitem[\protect\citeauthoryear{Kurucz}{1970}]{Kurucz1970} Kurucz R. L., 1970, Smithsonian Obs. Spec. Rep. No. 308.
\bibitem[\protect\citeauthoryear{Madau \&  Silk}{2005}]{Madau06}  Madau P., Silk J., 2005, MNRAS, 359, L37.
\bibitem[\protect\citeauthoryear{Marigo et al.}{2001}]{Marigo} Marigo P., Girardi L., Chiosi C., Wood P. R. 2001, \AAA, 371, 152.

\bibitem[\protect\citeauthoryear{Mayer \&  Duschl}{2005}]{Mayer}  Mayer M., Duschl W. J., 2005, MNRAS, 358, 614.
\bibitem[\protect\citeauthoryear{Michaud \& Proffitt}{1993}]{Michaud} Michaud G., Proffitt C. R., 1993, in Inside the stars, IAU Colloquium 137, eds Baglin A. \& Weiss W. W.,(ASP conference series, Vol 40), 246
\bibitem[\protect\citeauthoryear{Morel}{1997}]{Morel} Morel, P. 1997, \AAAS, 124, 597.
\bibitem[\protect\citeauthoryear{Nagakura \& Omukai}{2005}]{Nagakura} Nagakura T., Omukai K., 2005, MNRAS, 364, 1378.
\bibitem[\protect\citeauthoryear{Neale \& Tennyson}{1995}]{Neale95} Neale L., \& Tennyson J. 1995, \ApJ, 454, L169.
\bibitem[\protect\citeauthoryear{Rogers \& Nayfonov}{2002}]{OPALEoS}
Rogers, F. J., \& Nayfonov, A. 2002, ApJ, 576, 1064.
\bibitem[\protect\citeauthoryear{Saumon et al.}{1994}]{Saumon94}
Saumon, D., Bergeron, B., Lunine, J. I, Hubbard, W. B., \& Burrows, A., 1994, ApJ, 424, 333
\bibitem[\protect\citeauthoryear{Sauval \& Tatum}{1984}]{Sauval} Sauval A. J., \& Tatum J. B. 1984, \ApJS, 56, 193.
\bibitem[\protect\citeauthoryear{Siess, Livio \& Lattanzio}{Siess et al.}{2002}]{Siess} Siess, L., Livio, M., Lattanzio, J. 2002, \ApJ, 570, 329.
\bibitem[\protect\citeauthoryear{Stancil}{1996}]{Stancil} Stancil P., J., 1996, J. Quant. Spectrosc. Radiat. Transfer, 51, 655
\bibitem[\protect\citeauthoryear{Ryan et al.}{2000}]{Ryan} Ryan S. G., Beers T. C., Olive K. A., Fields B. D., Norris J. E., 2000, \ApJ, 530, L57
\bibitem[\protect\citeauthoryear{Tsujimoto \& Shigeyama}{2005}]{Tsujimoto} Tsujimoto T., Shigeyama T.,2005, \ApJ, 638, L109.
\bibitem[\protect\citeauthoryear{Wolniewicz, Simbotin \& Dalgarno}{Wolniewicz et al.}{1998}]{Wolniewicz} Wolniewicz, L., Simbotin, I., \& Dalgarno, A., 1998, \ApJS, 115, 293.


\end{thebibliography}
\end{document}